\documentclass[a4paper,11pt]{article}
\usepackage{latexsym}
\usepackage{cite}
\usepackage{epsfig}
\usepackage{amssymb}
\usepackage{latexsym}
\usepackage{epsfig, subfigure, subfloat, graphicx, float}
\usepackage{mathrsfs}
\usepackage{amsmath}
\usepackage{color}
\usepackage{mathtools}
\author{M. Khodaei\footnote{mahdi.khodaei@ut.ac.ir},  H. Mohseni Sadjadi\footnote{mohsenisad@ut.ac.ir}
\\ {\small Department of Physics, University of Tehran,}
\\ {\small P. O. B. 14395-547, Tehran 14399-55961, Iran}}
\title{No skyrmion hair for stationary spherically symmetric reflecting stars}
\begin{document}
\maketitle
 \begin{abstract}
We investigate the existence of the skyrmion field in the background of an asymptotically flat stationary reflecting star. For this purpose, we consider the Einstein-Skyrme system for which there is a skyrmion hair in the black hole case. We discuss spherically symmetric skyrmions and employ the hedgehog ansatz for the skyrmion field. We show that, in contrast to the black hole, there is no skyrmion hair for a reflecting star.
\end{abstract}

\section{Introduction}

According to the no-hair theorem\cite{nh}, all stationary black hole solutions of Einstein-Maxwell equations in general relativity are completely described by three observables: mass, electric charge, and angular momentum. This theorem states that no other quantum numbers can be observed by an outer observer when an object crosses the horizon. The difference between the electric charge, compared to other quantum numbers like the baryon number or the strangeness, comes from the Gauss law associated with the electric field.

Coupling of nonlinear fields to gravity has become one of the interesting subjects in recent years. In particular, it was shown that gravitationally bound configuration of nonabelian gauge fields in the Einstein-Yang-Mills theory can be one of the results of gravitational interactions \cite{Bartnik}. Studying of black hole solutions in these models shows that there may exist a nonlinear hair for a black hole\cite{Bizon, Garfinkle}, but they are unstable. Colored black hole solution (black hole solution of $SU(2)$ Einstein-Yang-Mills equations)\cite{Bizon, Volkov} was a counterexample for the no-hair theorem. This kind of solutions revealed that the no-hair theorem for the Einstein-Maxwell model, can not always be generalized to a more complicated system. But this solution is not stable against perturbations \cite{Strumann, Zhou}.

A black hole may have hair even in the absence of non-linear matter fields, e.g. in \cite{her2} a hairy stable black hole solution is found in the Einstein's gravity model with minimal coupling to a massive complex scalar field. This hair has not static limit and is supported by the rotation. The dynamical emergence of such a black hole from superradiance is studied in \cite{her3,her4}, and its stability (at least in a Hubble time) is verified in \cite{her5}.

One of the nonlinear models, which was developed to describe the baryons, is the Skyrme model \cite{skyrme}. In this model, baryons are topological solitons of a nonlinear meson field theory, and the winding number is identified as the baryon number. Stable topological soliton solution of the Einstein-Skyrme was studied in \cite{Bizon2}. The black hole solution of this model was considered in \cite{Droze}, and the linear stability of this solution was proved in \cite{Heusler}.

Recently, consideration of no-hair theorem near compact objects with a reflecting surface has become an interesting subject. A reflecting star is a physical compact object whose on the
surface the external field vanishes.  The non-existence of the scalar hair for a non-minimally coupled scalar field in the background of a reflecting star was proved in \cite{Hod}. Also in \cite{Hod 2}, the existence of a scalar hair for a charged massive scalar field in the background of a reflecting shell was verified. The no-hair theorem in the background of a reflecting star corresponding to the Dirichlet or the Neumann boundary condition was discussed also in \cite{peng1, peng2, Moh}.

The above-mentioned studies motivated us to investigate the possibility of the existence of skyrmion hair for a spherically symmetric reflecting star. In the following, we will prove that a spherically symmetric neutral reflecting star, in contrast to a black hole, cannot support a skyrmion field. The scheme of the paper is as follows:

In the second section, we briefly review the Skyrme model and in the third section, we study the solution of Einstein-Skyrme model in the background of a reflecting star and investigate the possibility to have non-trivial solutions.

\section{A review of the Skyrme model}
 The Skyrme Lagrangian is given by \cite{Manton,skyrme}
\begin{equation} \label{1}
\mathcal{L}_{S}=\frac{f_{\pi}^2}{16}g^{\mu\nu}tr(A_{\mu}A_{\nu})+\frac{1}{32a^2}g^{\rho\mu}g^{\sigma\nu}tr(F_{\mu\nu}F_{\rho\sigma}),
\end{equation}\label{2}
where $f_{\pi}$ is the pion decay constant and "a" is a dimensionless parameter.
In the above Lagrangian, $A_{\mu}$, and $F_{\mu\nu}$ are define as
\begin{equation}\label{2}
A_{\mu}=U^{\dagger}\partial_{\mu}U,\qquad F_{\mu\nu}=[A_{\mu},A_{\nu}].
\end{equation}
The field $U(x)$ is taken as a $SU(2)$ valued function. The first term in the right hand side of \eqref{1}, is the famous nonlinear sigma model Lagrangian and the second term is added by Skyrme in order to find topological soliton solutions. By deriving the canonical momentum, conjugated to the field $U(x)$
\begin{equation}\label{2.1}
\Pi_{ij}=\frac{\partial\mathcal{L}_{NL\sigma}}{\partial\dot{U_{ij}}},
\end{equation}
 one can find the Hamiltonian density:
\begin{equation}\label{2.2}
\mathcal{H}_{NL\sigma}=\Pi_{ij}\dot{U}_{ij}-\mathcal{L}_{NL\sigma}.
\end{equation}
The energy of the system will be defined by taking the space integral of the above Hamiltonian,
\begin{equation}\label{3}
E_{NL\sigma}=\int d^{3}x\mathcal{H}_{NL\sigma}\equiv E^{NL\sigma}_{rotation}+E^{NL\sigma}_{static},
\end{equation}
where
\begin{eqnarray}\label{4}
&&E^{NL\sigma}_{rotation}=\frac{f_{\pi}}{16}\int d^{3}x tr(\partial_{0}U^{\dagger}\partial_{0}U).\nonumber\\
&&E^{NL\sigma}_{static}=\frac{f_{\pi}}{16}\int d^{3}x tr(\partial_{i}U(x)\partial_{i}U(x)^{\dagger}).
\end{eqnarray}
 One can easily investigate the lack of topological soliton solution for the nonlinear sigma model.
By introducing the dimensionless variable $\tilde{x}$ which is defined by $x=\alpha\tilde{x}$, the static energy configuration becomes
\begin{equation}\label{5}
E_{NL\sigma}=\alpha\int d^{3}\tilde{x}\frac{F_{\pi}^2}{16}tr(U^{\dagger}\partial_{i}UU^{\dagger}\partial^{i}U).
\end{equation}
It is obvious that the integrand on the right hand side of the equation \eqref{5} is non-negative, and hence the energy is minimized at $\alpha =0$  with $E=0$. Therefore we just have a trivial solution. This is not the case when one considers the Skyrme term. The static energy of the Skyrme model is given by
\begin{equation}\label{6}
E^{Skyr}=\int d^{3}x \left[\frac{f_{\pi}^2}{16}tr(A_{i}A^{i})+\frac{1}{32a^2}tr(F_{ij}F^{ij})\right].
\end{equation}
By following the same arguments as above, one can check that this model can support nontrivial soliton solution. The static energy in terms of the rescaled variable $\tilde{x}$, is
\begin{equation}\label{7}
E_{static}^{Skyr}=\alpha\int d^{3}\tilde{x} \frac{f_{\pi}^2}{16}tr(A_{i}A^{i})+\frac{1}{\alpha}\int d^{3}\tilde{x}\frac{1}{32a^2}tr(F_{ij}F^{ij}).
\end{equation}
We see that the integrands on the right-hand side are non-negative therefore the energy is minimized at $\alpha\neq 0$.
Application of the following relation\cite{Y.L.Ma}
\begin{equation}\label{8}
tr(\epsilon^{ijk}A_{i}A_{j})^{2}=\frac{1}{2}tr[A_{i},A_{j}]^{2},
\end{equation}
in the equation \eqref{7}, leads to \cite{shiiki}
\begin{eqnarray}\label{9}
E_{static}^{Skyr}&=&\int d^{3}x tr\{\frac{f_{\pi}^2}{16}A_{i}A^{\dagger i}+\frac{1}{16a^2}(\epsilon_{ijk}A^{j}A^{k})(\epsilon^{ilm}A_{l}A_{m})^{\dagger}\}\nonumber\\
&=&E_{static}^{Skyr,(2)}+E_{static}^{Skyr(4)}.
\end{eqnarray}
The appropriate bounder condition, to obtain a finite $E$ is
\begin{equation}\label{9.1}
\qquad\qquad\qquad |X|\rightarrow\infty \Longrightarrow A_{i}\rightarrow 0,
\end{equation}
which means that $U$ approaches some constant matrix at infinity. Without loss of generality, we take this constant as the identity matrix
\begin{equation}\label{9.2}
\qquad\qquad\qquad |X|\rightarrow\infty \Longrightarrow U\rightarrow I.
\end{equation}
By application of the Cauchy-Schwartz inequality, $A^2 +B^2 \geq 2AB$, we have the following relation for the static energy
 \begin{eqnarray}\label{10}
 E_{static}^{Skyr}&=&\nonumber\\
 &&\quad\frac{f_{\pi}^{2}}{16}\int d^{3}x tr\left[A_{i}A^{i\dagger}+\frac{1}{a^{2}f_{\pi}^{2}}(\epsilon^{ijk}A_{j}A_{k})(\epsilon_{ilm}A^{l}A^{m})^{\dagger}\right]\nonumber\\
 &&\geq\frac{f_{\pi}^{2}}{16}\int d^{3}x\left| tr\left(\frac{2}{f_{\pi}a}\epsilon^{ijk}A_{i}A_{j}A_{k}\right)\right|\nonumber\\
 \end{eqnarray}
 We can see that the static energy is bounded from below, and the lower bound is greater than or equal to zero.
 The skyrmion's topological current is
 \begin{equation}\label{11}
B^{\mu}=-\frac{\epsilon^{\mu\nu\rho\sigma}}{24\pi^{2}}tr(A_{\nu}A_{\rho}A_{\sigma}).
\end{equation}
Zeroth component of this current is defined as a topological charge which in the Skyrme model is identified with the baryon number
\begin{equation}\label{12}
B=\int d^{3}xB^{0} = -\frac{1}{24\pi^{2}}\int d^{3}xtr(\epsilon_{ijk}A^{i}A^{j}A^{k}).
\end{equation}
In terms of topological charges, the relation \eqref{10} becomes
\begin{equation}\label{12.1}
 E_{static}^{Skyr}\geq\frac{3\pi^{2}f_{\pi}}{a}|B|,
\end{equation}
which is the Bogomol'ny bound. The lower limit is satisfied when
\begin{equation}\label{13}
A_{i}=\frac{1}{2af_{\pi}}\epsilon_{ijk}A^{j}A^{k},
\end{equation}
which means that $A_{i}$ is a self-dual field. Relation \eqref{13} is in contradiction with the Maurer-Cartan equation
\begin{equation}\label{14}
\partial_{\mu}A_{\nu}-\partial_{\nu}A_{\mu}=-[A_{\mu},A_{\nu}].
\end{equation}
We can interpret this incompatibility as: \textit{skyrmion energy should be larger than the Bogomol'ny bound}.
Now it is useful to show that the solutions of the Skyrme model are stable in three-dimensional space.
By following the earlier argument, by rescaling the field U(x) as $U(x)\rightarrow U(\alpha x)$,  we obtain
\begin{equation}\label{a-1}
E_{static}^{Skyr}(\alpha)=\alpha^{2-D}E_{static}^{Skyr,(2)}+\alpha^{4-D}E_{static}^{Skyr,(4)}.
\end{equation}
Therefore for three-dimensional space, we have\cite{Y.L.Ma}
\begin{eqnarray}
&&\frac{dE_{static}^{Skyr}(\alpha)}{d\alpha}\Big{|}_{\alpha =1}=-E_{static}^{Skyr,(2)}+E_{static}^{Skyr,(4)},\label{a-2}\\
&&\frac{d^{2}E_{static}^{Skyr}(\alpha)}{d\alpha^{2}}\Big{|}_{\alpha =1}=2E_{static}^{Skyr,(2)}.
\label{a-3}
\end{eqnarray}
The extremum stable condition
\begin{equation}
\frac{dE_{static}^{Skyr}(\alpha)}{d\alpha}\Big{|}_{\alpha =1}=0,
\end{equation}
leads to
\begin{equation}\label{a-4}
E_{static}^{Skyr,(2)}=E_{static}^{Skyr,(4)}=\frac{1}{2}E_{static}^{Skyr}.
\end{equation}
Now by using  \eqref{10}, one can see that $E_{static}^{Skyr,(2)}\geq 0$, therefore
\begin{equation}
\frac{d^{2}E_{static}^{Skyr}(\alpha)}{d\alpha^{2}}\Big{|}_{\alpha =1}=2E_{static}^{Skyr,(2)}\geq 0.
\end{equation}
This equation is the minimum stable condition for the Skyrme model solution.

\subsection*{The hedgehog ansatz}

By variation of the skyrmion action, the equation of motion is obtained as  \cite{Y.L.Ma,Manton}
\begin{equation}\label{16}
\frac{f_{\pi}^{2}}{16}(\partial_{\mu}A^{\mu})-\frac{1}{16a^2}\partial_{\mu}[A_{\nu},[A^{\mu},A^{\nu}]]=0.
\end{equation}
\eqref{16} is a nonlinear equation. To solve this equation, Skyrme proposed a so-called hedgehog ansatz under the assumption of maximally symmetric solution (see also  \cite{G.Holzwarth})
\begin{equation}
U(x)=exp(i\tau . \hat{X}\phi(r))=\cos\phi(r)+i\vec{n}.\tau\sin\phi(r),
\label{17}
\end{equation}
where $\vec{n}=X/r$ and $\tau$'s denote the Pauli matrices.
The boundary conditions are
\begin{equation}
\qquad \phi(0)=\pi ,\qquad\qquad \phi(\infty)=0.
\label{18}
\end{equation}
 Inserting the hedgehog ansatz into the energy relation \eqref{6} leads to
\begin{eqnarray}\label{19}
E_{static}&=&4\pi\int_{0}^{\infty}drr^{2}[\frac{f_{\pi}^{2}}{8}\left(\phi^{\prime 2}+\frac{2 \sin^{2}\phi}{r^{2}}\right)\nonumber\\
&&+\frac{\sin^{2}\phi}{2a^{2}r^{2}}\left(2\phi^{\prime 2}+\frac{\sin^{2}\phi}{r^{2}}\right)].
\end{eqnarray}
By varying the above equation with respect to the function $\phi $,  one can calculate the extremum of the energy
\begin{equation}\label{20}
\left(\frac{x^{2}}{4}+2\sin^{2}\phi\right)\phi^{\prime\prime}+\frac{x}{2}\phi^{\prime}+\phi^{\prime 2}\sin 2\phi-\frac{\sin^{2}\phi\sin 2\phi}{x^{2}}=0.
\end{equation}
We use new the dimensionless variable $x=af_{\pi}r$ in relation \eqref{20}.

\section{The Einstein-Skyrme model in the background of a spherically symmetric reflecting star}
In this section, we want to consider the Einstein-Skyrme model \cite{luck} in the background of a reflecting star whose the reflecting surface is located out of the horizon.
The Einstein-Skyrme system is defined by the lagrangian
\begin{eqnarray}
\mathcal{L} &=&\mathcal{L}_{G}+\mathcal{L}_{S}
\nonumber\\
&=&\frac{R}{16\pi G}+\frac{f^{2}_{\pi}}{16}g^{\mu\nu}tr(A_{\mu}A_{\nu})+\frac{1}{32a^{2}}g^{\rho\mu}g^{\sigma\nu}tr(F_{\mu\nu}F_{\rho\sigma}),
\label{21}
\end{eqnarray}
where R is the Ricci scalar and G is the Newton's gravitational constant. The action of the Einstein-Skyrme system is given by
\begin{eqnarray}
S&=&S_{G}+S_{S}\nonumber\\
&=&\int d^{4}x\sqrt{-g}(\mathcal{L}_{G}+\mathcal{L}_{S}).
\label{22}
\end{eqnarray}
Variation with respect to the metric leads to
\begin{equation}
G_{\mu\nu}=8\pi GT_{\mu\nu},
\label{23}
\end{equation}
where $T_{\mu\nu}$ denotes energy-momentum tensor.  One can prove that\cite{Droze}
\begin{eqnarray}\label{24}
T_{\mu\nu}&=&-g_{\mu\nu}L_{S}+2\frac{\partial L_{S}}{\partial g^{\mu\nu}}\nonumber\\
&=&\frac{1}{8}f_{\pi}^{2}tr(A_{\mu}A_{\nu}-\frac{1}{2}g_{\mu\nu}A_{\alpha}A^{\alpha})\nonumber\\
\quad &&+\frac{1}{8a^{2}}tr(F_{\mu\alpha}F_{\nu\beta}g^{\alpha\beta}-\frac{1}{4}g_{\mu\nu}F_{\alpha\beta}F^{\alpha\beta}),
\end{eqnarray}
where $L_{s}=\mathcal{L}_{S}\sqrt{-g}$.
We are looking for static spherical symmetric solutions. We use the hedgehog ansatz for the field U
\begin{equation}
U(r)=\cos \phi(r)+i\vec{n}.\vec{\tau}\sin \phi(r).
\label{25}
\end{equation}
We choose a spherical symmetric metric\cite{shiiki} as
\begin{equation}\label{26}
ds^{2}=-N(r)C(r)dt^{2}+\dfrac{1}{C(r)}dr^{2}+r^{2}d\Omega ^{2},
\end{equation}
where
\begin{equation*}
C(r)=1-\dfrac{2GM}{r}.
\end{equation*}
The black-hole admits skyrmion hair. Let us investigate, what happens for a reflecting star. Since the reflecting surface is out of the horizon, we expect
\begin{equation*}
C(r)>0\qquad for\qquad r_{s}\leq r\leq \infty .
\end{equation*}

The static energy for the chiral field is given by\cite{shiiki}
\begin{eqnarray}
E_{static}&=&\dfrac{4\pi f_{\pi}}{a}\int _{x_{s}}^{\infty}\Big[\frac{1}{8}\left(C\phi^{\prime 2}+\dfrac{2 \sin^{2}\phi}{x^{2}}\right)\nonumber\\
&&+\frac{\sin^{2}\phi}{x^{2}}\left(2C\phi^{\prime 2}+\dfrac{\sin^{2}\phi}{x^{2}}\right)\Big] x^{2} dx,
\label{28}
\end{eqnarray}
where we have used the dimensionless variable $x=af_{\pi}r$. Note that, our framework is the outer side of the reflecting surface and therefore the integral is from the reflecting surface to the infinity. Now in the presence of gravity, the topological current\eqref{11} is defined as
\begin{equation}
B^{\mu}=-\frac{\epsilon^{\mu\nu\rho\sigma}}{24\pi ^{2}}\frac{1}{\sqrt{-g}}tr\left(A_{\nu}A_{\rho}A_{\sigma}\right).
\end{equation}
As previously noted, the zeroth component of the topological current is defined as the baryon number density
\begin{equation}
B^{0}=-\frac{1}{2\pi ^{2}}\frac{1}{N}\frac{\phi^{\prime}\sin ^{2}\phi}{r^{2}}.
\end{equation}
Not let us put Neumann boundary condition on $U$, such that
\begin{equation}
U^{\prime}\Big{|}_{r_{s}}=0,\,\,\, U(\infty)=constant.
\label{27}
\end{equation}
Note that in the hedgehog ansatz , $U\neq 0$. This is equivalent to take the following boundary condition for the profile function
\begin{equation}
\frac{d\phi(x)}{dx}\Big{|}_{x=x_{s}}=0
\end{equation}
and also for the other boundary we take $\phi(\infty)=0$ giving $U(\infty)=I$. In this situation the baryon number becomes
\begin{equation}
B=\int\sqrt{-g}B^{0}d^{3}x=-\dfrac{2}{\pi}\int_{\phi_{s}}^{0}\sin ^{2}\phi d\phi=\dfrac{1}{2\pi}(2\phi_{s}-\sin 2\phi_{s}).
\label{29}
\end{equation}
 This shows that the amount of the baryon number depends on the profile function on the reflecting surface.

From Einstein equation \eqref{23} and the metric ansatz \eqref{26} we will have\cite{shiiki}
 \begin{equation}
N^{\prime}=\dfrac{\alpha}{4}\left(x+\dfrac{8\sin ^{2}\phi}{x}\right)N\phi^{\prime 2},
\label{30}
\end{equation}
 and
 \begin{equation}
\dfrac{\alpha}{8}\left[(x^{2}+8 \sin ^{2}\phi)C\phi^{\prime 2}+2\sin ^{2}\phi+\dfrac{4\sin ^{4}\phi}{x^{2}}\right]=0,
\label{31}
\end{equation}
 where the coupling constant $\alpha$ is defined as
 \begin{equation*}
  \alpha=4\pi Gf_{\pi}^{2}.
 \end{equation*}
 Taking the variation of the static energy\eqref{28} with respect to $\phi(x)$, yields
 \begin{eqnarray}
\phi^{\prime\prime}&=&\dfrac{1}{NC(x^{2}+8\sin ^{2}\phi)}[-(x^{2}+8\sin ^{2}\phi)N^{\prime}C\phi^{\prime}\nonumber\\
&&+\left(1+\dfrac{4\sin ^{2}\phi}{x^{2}}+4C\phi^{\prime 2}\right) N\sin 2\phi \nonumber\\
&&-2(x+4 \phi^{\prime}\sin 2\phi)NC\phi^{\prime}-2\left(1+\dfrac{8\sin ^{2}\phi}{x^{2}}\right)MN\phi^{\prime}] .
\label{32}
\end{eqnarray}
 Equation \eqref{30} on the reflecting surface yields
 \begin{equation}
\dfrac{dN(x)}{dx}\Big{|}_{x=x_{s}}=0.
\label{33}
\end{equation}
  substitution  \eqref{33} into \eqref{31} gives
 \begin{equation}
2\sin^{2}(\phi_{s})+\dfrac{4}{x^{2}_{s}}\sin^{4}(\phi_{s})=0\longrightarrow \phi_{s}=n\pi ,\quad n=0,1,2,... .
\label{34}
\end{equation}
By applying equations \eqref{33}, \eqref{34} in \eqref{31}, we obtain
\begin{equation}
\dfrac{d^{2}\phi(x)}{dx^{2}}\Big{|}_{x=x_{s}}=0.
\label{35}
\end{equation}
 Similarly, by substituting  \eqref{33}, \eqref{34}, and \eqref{35} into the derivatives of \eqref{30}, and \eqref{31}, we see that all the higher-order derivatives of the profile function on the reflecting surface are zero. We can expand the analytic continuous profile function around the reflecting surface as
\begin{eqnarray}
\phi(x)&=& \phi(x_{s})+\phi^{\prime}(x)\Big{|}_{x=x_{s}}(x-x_{s})\nonumber\\
&&+\dfrac{1}{2}\phi^{\prime\prime}(x)\Big{|}_{x=x_{s}}(x-x_{s})^2+... .
\end{eqnarray}
so we find that $\phi$ is a trivial constant. If we take $\phi(\infty)=0$, this constant is zero. From (\ref{29})we find  $B=0$. So, in contrast to the black hole, there is no skyrmion hair for the corresponding reflecting star.

At the end let us note that if we imposed Dirichlet boundary condition on $\phi(x_{s})=0$, we would get the same result. To see this, let us take $\phi(x_s)=0$. From \eqref{31} we have
 \begin{equation}
\phi(x_{s})=0\longrightarrow \dfrac{d\phi(x)}{dx}\Big{|}_{x=x_{s}}=0.
\label{37}
\end{equation}
Using this result in \eqref{30} yields
\begin{equation}
\dfrac{dN(x)}{dx}\Big{|}_{x=x_{s}}=0
\label{38}
\end{equation}
Using \eqref{37},\eqref{38} in \eqref{32} yields
\begin{equation}
\dfrac{d^{2}\phi(x)}{dx^{2}}\Big{|}_{x=x_{s}}=0.
\label{39}
\end{equation}
By consecutive derivation of (\ref{30}) and (\ref{32}) we find
\begin{equation}
\dfrac{d^{n}\phi(x)}{dx^{n}}\Big{|}_{x=x_{s}}=0 ,  \qquad  n=0,1,2,3,...,
\end{equation}
and
\begin{equation}
\dfrac{d^{m}N(x)}{dx^{m}}|_{x=x_{s}}=0,  \qquad m=1,2,3,... .
\end{equation}

 We see that all of the profile function's and also all the metric's coefficient (N)  derivatives on the reflecting surface vanish. Therefore the only solution for \eqref{32} is the trivial solution and the metric outside of the star is the Schwarzschild metric.

Existence of skyrmion hair was investigated in \cite{luck}, and there was shown that we may have hairy black holes in the Einstein-Skyrme model. If we put the reflecting surface at a (possible) zero (or an extremum) of the scalar profile, then based on our above argument we will have a trivial profile function outside the reflecting surface. Note that $r_s>r_h$, where $r_h$ is the horizon radius.\newpage

{\bf{Conclusion}}

After a brief review of the skyrmion model, which is derived by adding the Skyrme term to the non-linear sigma model, we studied if topological stable soliton solutions can exist in the background of a spherically symmetric reflecting star. We showed that in contrast to the black hole \cite{dv}, such a hair does not exist and we cannot assign a Baryon number to the reflecting star.

 \end{document}